\documentclass[utf8]{frontiersSCNS} 

\usepackage{url,hyperref,lineno,microtype,subcaption}
\usepackage[onehalfspacing]{setspace}
\usepackage{color}
\usepackage{colortbl}
\usepackage{booktabs}

\def\keyFont{\fontsize{8}{11}\helveticabold }
\def\firstAuthorLast{Oliveira e Silva {et~al.}} 
\def\Authors{Alexandre Jos\'e de Oliveira e Silva\,$^{1,2,3}$, Caius L. Selhorst\,$^{1,*}$, Joaquim E. R. Costa\,$^{3}$, Paulo J.~ A.~Sim\~oes\,$^{4,5}$, C. Guillermo Gim{\'e}nez de Castro\,$^{4,6}$, Sven Wedemeyer\,$^{7}$, Stephen M. White\,$^{8}$, Roman Braj\v sa\,$^{9}$ and Adriana Valio\,$^4$}
\def\Address{$^{1}$ NAT - N\'ucleo de Astrof\'isica, Universidade Cidade de S\~ao Paulo, S\~ao Paulo, SP, Brazil\\
$^{2}$IP\&D - Instituto de Pesquisa e Desenvolvimento, Universidade do Vale do Para\'iba, S\~ao Jos\'e dos Campos, SP, Brazil \\
$^{3}$INPE - Instituto Nacional de Pesquisas Espaciais (DICEP), S\~ao Jos\'e dos Campos, SP,  Brazil \\
$^{4}$Centro de R\'adio Astronomia e Astrof\'isica Mackenzie, Escola de Engenharia, Universidade Presbiteriana Mackenzie,
Pr\'edio 45 T, cobertura, R. da Consolação, 896--7o andar--Consolação,
São Paulo, Brazil\\
$^{5}$SUPA School of Physics and Astronomy, University of Glasgow, Glasgow G128QQ, UK \\
$^{6}$Instituto de Astronom{\'\i}a y F{\'\i}sica del Espacio, CONICET-UBA, Ciudad Universitaria\\ Buenos Aires, Argentina\\
$^{7}$ Rosseland Centre for Solar Physics, Institute of Theoretical Astrophysics, University of Oslo, Postboks 1029 Blindern, N-0315 Oslo, Norway\\
$^{8}$ Space Vehicles Directorate, Air Force Research Laboratory, Albuquerque, NM, USA\\
$^{9}$ Hvar Observatory, Faculty of Geodesy, University of Zagreb, Zagreb, Croatia
}

\def\corrAuthor{Caius L. Selhorst}

\def\corrEmail{caiuslucius@gmail.com}

\begin{document}
\onecolumn
\firstpage{1}



\title{A Genetic Algorithm to model Solar Radio Active Regions from 3D Magnetic Field Extrapolations}



\author[\firstAuthorLast ]{\Authors} 
\address{} 
\correspondance{} 

\extraAuth{}

\newcommand{\ps}[1]{{\color{red}{#1}}}

\maketitle

\begin{abstract}

\section{}
In recent decades our understanding of solar active regions (ARs) has improved substantially due to observations made with better angular resolution and wider spectral coverage. While prior AR observations have shown that these structures were always brighter than the quiet Sun at centimeter wavelengths, recent observations at millimeter and submillimeter wavelengths have shown ARs with well defined dark umbrae. Given this new information, it is now necessary to update our understanding and models of the solar atmosphere in active regions. In this work, we present a data-constrained model of the AR solar atmosphere, in which we use brightness temperature measurements of NOAA 12470 at three radio frequencies: 17, 100 and 230~GHz. The observations at 17~GHz were made by the Nobeyama Radioheliograph (NoRH), while the observations at 100 and 230~GHz were obtained by the Atacama Large Millimeter/submillimeter Array (ALMA). Based on our model, which assumes that the radio emission originates from thermal free-free and gyroresonance processes, we  calculate radio brightness temperature maps that can be compared with the observations. The magnetic field at distinct atmospheric heights was determined in our modelling process by force-free field extrapolation using photospheric magnetograms taken by the  Helioseismic and Magnetic Imager (HMI) on board the Solar Dynamics Observatory (SDO). In order to determine the best plasma temperature and density height profiles necessary to match the observations, the model uses a genetic algorithm that modifies a standard quiet Sun atmospheric model. Our results show that the height of the transition region (TR) of the modelled atmosphere varies with the type of region being modelled: for umbrae the TR is located at $1080\pm20$~km above the solar surface; for penumbrae, the TR is located at $1800\pm50$~km; and for bright regions outside sunspots, the TR is located at $2000\pm100$~km. With these results, we find good agreement  with the observed AR brightness temperature maps. Our modelled AR can be used to estimate the emission at frequencies without observational coverage. 

\tiny
 \keyFont{ \section{Keywords:} Sun: radio radiation, Sun: atmosphere,  Sun: magnetic fields, Force-free field extrapolation} 
\end{abstract}

\section{Introduction}
Numerous semi-empirical models of the solar atmosphere are intended to reproduce solar active region (AR) observations with success in a specific wavelength range. Most of these efforts have been based on optical and UV line observations \citep[e.g.: ][]{VAL1981,FAL1993,Fontenla1999,Fontenla2009} and have had success modelling the observations at these wavelengths. Nevertheless, these models are less successful in reproducing radio observations of ARs, perhaps due to the presence of distinct emission mechanisms at radio wavelengths and the lack of comparable spatial resolution.     

In the last decades our knowledge of solar active regions has improved substantially due to observations made with better angular resolution. Recent observations at millimeter and submillimeter wavelengths have shown ARs with well defined dark umbrae \citep{Loukitcheva2014,Iwai2015,Shimojo2017}. Moreover, \cite{Iwai2016} suggested that even at radio frequencies as low as 34~GHz the observed brightness temperature ( $T_\mathrm{b}$) of the umbral region is almost the same as that of the quiet region, which indicates that the height (and therefore temperature) in the atmosphere at which its emission becomes optically thick should be lower than that predicted by the models. 

Great advances in the understanding of AR behavior have been acquired with the daily NoRH \citep{Nakajima1994} observations at 17~GHz since 1992, and also 34~GHz after 1996. \cite{Vourlidas2006} studied 529 ARs observed by the NoRH at 17~GHz (1992--1994), and concluded that the ARs with polarization greater than $30\%$ contain a gyroresonance core that increases their brightness temperatures to $T_\mathrm{b}\gtrsim 10^5$~K, well above the quiet Sun brightness temperature ( $T_\mathrm{b,QS}$) of $10000$~K at 17 GHz. Moreover, they also argued that these high $T_\mathrm{b}$ values were due to opacity at the gyroresonance 3rd harmonic, requiring 2000~G of magnetic field intensity ($|\vec{B}|$) in the corona, and therefore at least $|\vec{B}|=2200$~G at the photospheric level. 

At submillimeter wavelenghts, \cite{Silva2005} analyzed a total of 23 ARs observed during 2002 at 212 and 405~GHz from the Submillimeter Solar Telescope \citep[SST, ][]{Kaufmann2008} and combined with maps at 17 and 34~GHz from NoRH. The flux density spectra at these frequencies was found to increase with frequency with a slope of 2, indicating that the emission from these active regions is predominantly due to thermal bremsstrahlung. Moreover, \cite{Valle2021} made a similar analyses with the inclusion of the ALMA single-dish data reaching the same conclusions. 

Further advances in the study of ARs became possible in 2016, with the start of ALMA observations at millimeter/submillimeter wavelengths \citep{Wedemeyer2016}. During the Science Verification period, 2015 December 16–20, AR 12470, including a sunspot umbra, was observed by the ALMA interferometric array at Band 3 (84--116 GHz) and Band 6 (211–275 GHz). Whereas the sunspot umbra exhibited a continuous dark region at Band 6 \citep{Shimojo2017}, at Band 3 the center of the umbra showed a bright structure with $T_\mathrm{b}$ 800~K above the dark region around it \citep{Iwai2017}. This Band 3 brightness enhancement may be an intrinsic feature of the sunspot umbra at chromospheric heights, such as a manifestation of umbral flashes, or it could be related to a coronal plume.

\cite{Selhorst2008} were able to reproduce the brightness temperatures and the spatial structure of AR NOAA 10008 seen in NoRH observations at 17 and 34~GHz. That work used photospheric magnetic field extrapolation to derive $\vec{B}$ in the atmosphere\footnote{The 3D magnetic field extrapolation used in \cite{Selhorst2008} has been integrated in the fast algorithm GX Simulator tool available in SolarSoft \citep{Nita2015,Nita2018}} and suggested that the AR chromospheric temperature and density gradients were steeper than those proposed for the quiet Sun. Also, the temperature and density in the corona needed to be greater than expected in the quiet Sun, as expected. With these assumptions, the model was able to reproduce the high $T_\mathrm{b}$ observed in the 17~GHz gyroresonance core, and also the free-free emission observed in non-polarized areas at both 17 and 34~GHz.

\cite{Brajsa2009} studied active regions observed at 37 GHz with the 14-m antenna of the Mets\"ahovi Radio Observatory and compared the measured intensities with radiation models. They concluded that thermal bremsstrahlung can explain the observed radiation of ARs, while thermal gyromagnetic emission can, with high probability, be excluded as a possible radiation mechanism at the frequency considered (37 GHz). Further, \cite{Brajsa2018} compared intensities measured in full-disc solar ALMA maps taken at 248 GHz with model-based prediction of the brightness temperatures. Again, they concluded that the thermal bremsstrahlung is the main radiation mechanism responsible for the AR emission, also at this observing frequency (248 GHz). 

As an improvement of the AR modelling presented in \cite{Selhorst2008,Selhorst2009}, in this work we apply a genetic algorithm \citep[GA; ][]{Charbonneau1995} to modify a standard quiet Sun atmospheric model in order to determine the best plasma temperature and density height profiles necessary to match the observations of NOAA 12470 obtained by ALMA (single-dish maps at 100 and 230~GHz) and NoRH (17~GHz interferometric map).  

Finally, we note that \cite{Brajsa2020} performed a preliminary analysis of the magnetic structure above the same AR 12470 using LOS photospheric magnetograms and a potential-field source surface (PFSS) model to extrapolate magnetic fields into the solar chromosphere and corona. Results of the model were compared with the ALMA single-dish (248 GHz) and interferometric (100 GHz) measurements of the same AR. The general extrapolated magnetic structure is consistent with the ALMA observations, but a detailed analysis and comparison with ALMA small-scale features was not possible with the model used and requires a more detailed magnetic field extrapolation model, what is performed in the present work.

\section{Atmospheric modelling via genetic algorithm}\label{sec:models}

The purpose of this study is to estimate the temperature and electron density as a function of height in the solar atmosphere, as constrained by brightness temperature $T_\mathrm{b}$ maps of radio observations. The different frequencies provide access to different layers of the atmosphere due to the frequency dependence of opacity. An initial 1D atmospheric model is used as a seed for the GA: at each step, $T_\mathrm{b}$ is obtained via the calculation of radiative transfer incorporating opacity from both the thermal bremsstrahlung and gyroresonance emission mechanisms. For these calculation, the 3D magnetic field structure above the AR is derived from linear force-free extrapolations (LFF) from photospheric magnetograms. The GA iteratively updates the atmospheric plasma trapped by magnetic field lines until a best model is found via $\chi^2$ minimization of the calculated and observed $T_\mathrm{b}$ values. This process is run independently for each pixel of the observed radio maps, resulting in an atmospheric model ($T$ and $n$ versus height) for each line of sight. Together, the results for all pixels provide the 3D atmospheric model for the active region.

\subsection{Genetic Algorithm Features}

In order to determine the best plasma temperature and density height profiles necessary to match the observations, we employ the genetic algorithm Pikaia \citep{Charbonneau1995}. 

The atmospheric model developed by \citet[][hereafter SSC]{Selhorst2005b} is used as the seed in the genetic algorithm, that changes the seed model to find the best fit for the observational measurements. Five free parameters are supplied to the GA representing the atmospheric model. These parameters are based on the ideas presented in \cite{Selhorst2008}, as follows:

\begin{itemize}
    \item $\nabla T$ -- changes the AR temperature gradient in the chromosphere;
    \item $\nabla n_e$ -- changes the AR electron density gradient in the chromosphere;
    \item $N_T$ -- changes the AR coronal temperature by a constant value, in which $N_T=\frac{T_{AR}}{T_{qS}}$;
    \item $N_{n_e}$ -- changes the AR coronal electron density by a constant value, in which $N_{n_e}=\frac{n_{e_{AR}}}{n_{e_{qS}}}$;
    \item $\Delta h$ -- changes the TR position. 
\end{itemize}

These five free parameters combined build new electron density and temperature profiles, which are used, along with the magnetic field values for the line-of-sight column, to calculate the bremsstrahlung and gyroresonance absorption coefficients. The radiative transfer is then performed and the resulting $T_\mathrm{b}$ values are compared with the observations. {Nevertheless, the GA used here is a mathematical solution and it does not verify if the atmospheric plasma is in local thermodynamic equilibrium (LTE) or not.}

A generation of child atmospheric models is created by the algorithm based on the parent models with the lowest $\chi^2$ values, using the genetic algorithm described in \cite{Charbonneau1995}. The GA starts with the choice of minimum and maximum values for each one of the five free parameters. The method used here was full-generation replacement using ten generations with one hundred children each. The crossover probability was 0.85 and the mutation mode was variable. The initial mutation rate, that is, the initial probability that any gene locus will mutate in any generation, was 0.005. Finally, the mutation rate range was between 0.0005 and 0.25. In our runs the $\chi^2$ measure typically stabilizes to a minimum value by about the seventh generation.

\subsection{Magnetic field extrapolation}\label{sec:Extrapolation}

Photospheric magnetograms are used as the boundary condition for an extrapolation of the field above the photosphere using the linear force-free field approximation. The fundamental equation that describes a force-free field (assuming that the current density is parallel to the magnetic field at any point in space) is

\begin{equation}
   \vec{\nabla} \times \vec{B}=\alpha\vec{B}\\ 
   \label{alfa2}
\end{equation}

\noindent where $\alpha$ is a proportionality function and represents the current density distribution, called a force-free function.

The extrapolation routine is based on the works of \cite{Nakagawa1972} and \cite{Seehafer1978} and was previously implemented in IDL by J. E. R. Costa and T. S. N. Pinto \citep[used in ][]{Selhorst2005d,Selhorst2008,Nita2018}. The calculation starts from the line-of-sight magnetic field component at the photosphere and determines the magnetic field above it for both the potential field ($\alpha=0$) and the force-free field ($\alpha\neq0$). The result is three cubes of intensities, one for each component of the magnetic field vector ($B_x$, $B_y$, and $B_z$) and another data cube for the positions of the magnetic field lines. The positions of the magnetic field lines determine where the atmospheric plasma is changed with respect to the quiet Sun. Moreover, the $x$ and $y$ dimensions of the cubes have the same dimensions as the selected active region area. In this work, the extrapolated field lines were calculated only above positions with $|\vec{B}|_{min}=200$~G at the photosphere. A different minimum value could be chosen by the user, but we found that the inclusion of magnetic fields below this limit resulted in the simulated ARs appearing much larger than the observed ARs. This approach in effect determines one parameter, i.e., the lower limit of the field strength needed to best reproduce the observations.

\subsection{Emission mechanisms}\label{sec:emission}

The brightness temperature $T_\mathrm{b}$ is obtained by computing the radiative transfer along the line of sight for each column of the AR. 
At radio frequencies, in the quiescent solar atmosphere, the main emission mechanisms are bremsstrahlung and gyroresonance \citep{Dulk1985}.
The thermal bremsstrahlung absorption coefficient ($\kappa_b$) is calculated by
\begin{equation}\label{bremsstrahlung}
   \kappa_b \approx 9.78\times10^{-3} \frac{n_e}{\nu^2~T^{3/2}}\sum_i Z_i^2~n_i\times g(T,\nu) 
\end{equation}
\noindent where $\nu$ is the observed frequency (Hz), $T$ is the plasma temperature and $n_e$ is the electron density. Only the contributions of collisions with protons was considered, since its density is much larger than that of other ion species ($n_i$). 

{Following \cite{Zirin1988}, it was assumed $Z=1.178$, if a fully hydrogen atmosphere was considered, i.e. $Z=1$,~the resulting $T_\mathrm{b}$ should be only $\sim 3\%$ smaller for all frequencies simulated here}. 

The Gaunt factor (g(T,$\nu$)) used is as follow:

\begin{align*}\label{gaunt}
   g(T,\nu)=\left\{\begin{array}{ll}
   18.2+\ln{T^{3/2}} - \ln \nu &(T < 2\times10^5 \text{~K}) \\
   24.5+\ln T - \ln \nu        &(T > 2\times10^5 \text{~K}) 
   \end{array}\right.
\end{align*} 

The gyroresonance absorption coefficient  ($\kappa_g$) is 

\begin{eqnarray} \label{emiss_giro3}
   \kappa_g(s,\theta) &=& \int_{-\infty}^{\infty} \kappa_g(s,\theta)~\frac{d\nu}{\nu_b} \nonumber\\
   &=& \left(\frac{\pi}{2}\right)^\frac{5}{2}~\frac{2}{c}~\frac{\nu_p^2}{\nu}~\frac{s^2}{s!} \left(\frac{s^2~\beta_o^2~\sin^2\theta}{2}\right)^{s-1} \times(1-\sigma|\text{cos}~\theta|)^2 
\end{eqnarray}

\noindent where $\sigma=+1$ for the o-mode and $\sigma=-1$ for the x-mode, $\beta_o^2=kT/mc^2$. The harmonic is defined as $s=\nu/\nu_B$, where $\nu_B$ is electron-cyclotron frequency and $\nu_p$ is the electron plasma frequency:

\begin{eqnarray}
\nu_B &\approx& 2.8\times 10^6 B \nonumber\\
\nu_p &\approx& 9000~ n_e^{1/2} \nonumber
\end{eqnarray}

Since the contribution of higher harmonics to the radio emission is very small \citep{Shibasaki1994}, the gyroresonance emission was only calculated for harmonics smaller than 5 ($s\leq 5$).

Following \cite{Zirin1988}, radiative transfer yields the brightness temperature as a function of wavelength as:
\begin{equation}\label{tb_transf}
   T_b(\nu) = \int T\kappa_\nu~e^{-\tau_\nu} dL
\end{equation}

\noindent where $\kappa_\nu = \kappa_b+\kappa_g$, $dL$ is the distance element towards the observer, and $\tau_\nu$ is the optical depth

\begin{equation}\label{deftau}
   \tau_\nu = \int \kappa_\nu~dL 
\end{equation}


\section{Observations} \label{sec:observation}

In this work, we selected the active region AR NOAA 12470 to test our algorithm. This region was selected due to the availability of radio observations at three frequencies: 17 GHz, from the Nobeyama Radio Heliograph (NoRH), 100 and 230 GHz from the Atacama Large Millimetric and submillimetric Array (ALMA) \citep{Shimojo2017,Valle2021}. 

{Here, the LFF extrapolations algorithm used the line-of-sight (LOS) magnetogram obtained by the Helioseismic and Magnetic Imager \citep[HMI, ][]{Schou2012} on the Solar Dynamics Observatory \citep[SDO, ][]{Pesnell2012}, nevertheless, the LFF extrapolations could also be applied to vector magnetograms.} HMI operates at wavelengths around 6173 \AA (Fe I) and provides LOS magnetograms of $4096\times4096$ pixels with $\sim0.5$~arcsec spatial resolution.

\subsection{Magnetogram} \label{sec:magneto}

\begin{figure}[ht!]
\begin{center}
\includegraphics[width=15cm]{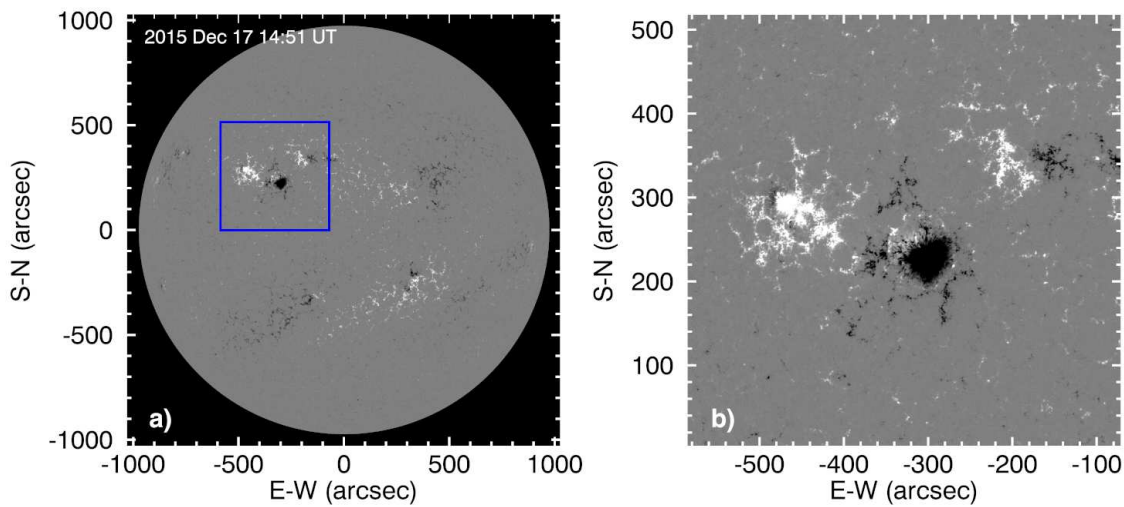}
\end{center}
\caption{The left image shows the SDO/HMI LOS magnetogram obtained on 2015 Dec 17 at 14:51 UT. The blue rectangle indicates the selected area of active region AR12470, shown in expanded form in the right panel. The region shown is $516.4''\times516.4''$, centered at $17.40^\circ$N,$17.68^\circ$E.}
\label{fig:magneto}
\end{figure}

Figure~\ref{fig:magneto} (left panel) shows the HMI line of sight (LOS) magnetogram obtained on December 17, 2015 at 14:51 UT. This full Sun image has an apparent radius of 975'' and magnetic field intensities ($|\vec{B}|$) up to 2300~G. The active region is shown in detail in Figure~\ref{fig:magneto}, right panel, centered in the point of greatest magnetic field intensity ($|\vec{B}|_{max}$). The selected area included the whole AR NOAA 12470, and also part of the AR NOAA 12469 located to the northwest of the central umbra.

The line-of-sight magnetogram will underestimate field strengths in AR 12470 due to projection effects, since the region is significantly offset from disk center. To correct for this, the AR has been rotated to solar disk center, solar spherical curvature has been removed, and the magnetic field intensities have been corrected for projection. This increases the maximum field strength from $2300$~G to $\sim2500$~G.

\subsection{Radio maps} \label{sec:radio}
 
The ALMA maps used here were obtained during the Science Verification period, 2015 December 16–20 \footnote{https://almascience.eso.org/alma-data/science-verification}. The maps were obtained by the fast-scan single-dish method
described in \cite{White2017}, at Band 3 (84–116 GHz) and Band 6 (211–275 GHz). Following \cite{Selhorst2019}, we took 100 and 230~GHz as the reference frequencies for Bands 3 and 6, respectively. The maps were made from observations of a 12m diameter antenna, resulting in nominal spatial resolutions of 25'' and 58'' at 230 and 100~GHz, respectively.

\begin{figure}[ht!]
\begin{center}
\includegraphics[width=17cm]{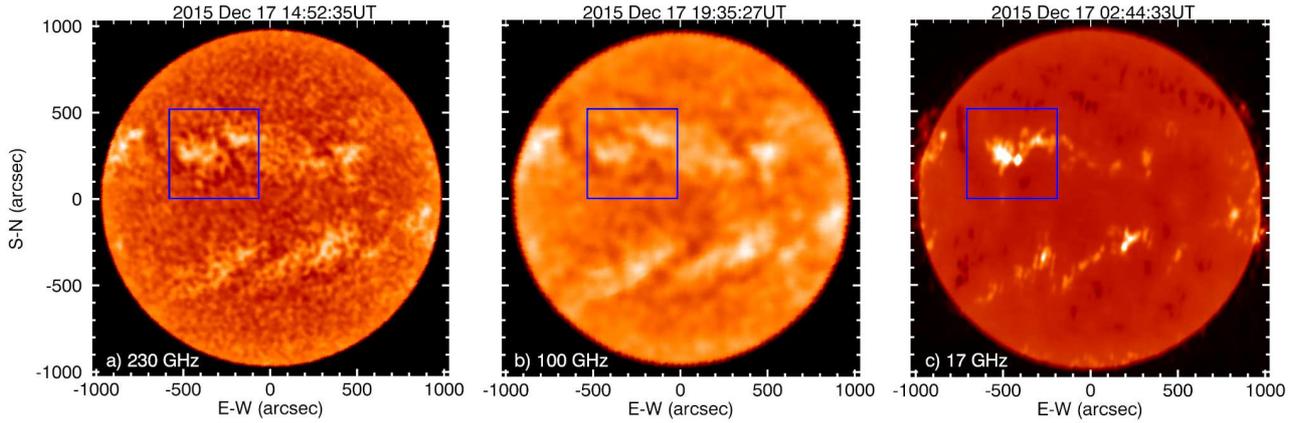}
\end{center}
\caption{Full disc solar maps obtained at (a) 230~GHz (a), (b) 100~GHz and (c) the NoRH map at 17~GHz. The blue rectangle in each panel indicates the selected area of the active region AR12470, taking into account solar rotation given timing differences relative to the HMI magnetogram. 
}
\label{fig:mapradio}
\end{figure}

Since the radio maps were not obtained at the same time as the magnetogram, it was necessary to take into account solar rotation when aligning the radio data with the magnetogram. The nominal time of the 230~GHz map is 14:52 UT, i.e., only 1 minute difference from the nominal time of the HMI observation, requiring just $-0.01^\circ$ of longitudinal rotation. The nominal time of the 100~GHz map was  19:35 UT, requiring a longitudinal rotation of $-2.85^\circ$. Moreover, the ALMA maps come in distinct sizes, $800\times800$ pixels$^2$ (3'' pixel) at 230~GHz and $400\times400$ pixels$^2$ (6'' pixel) at 100~GHz. Thus for a better comparison with the model results they were resized to match the HMI magnetogram  ($4096\times4096$ pixels$^2$).   

While the quiet Sun brightness temperature ( $T_\mathrm{b,QS}$) at 230~GHz is $6210\pm110$~K \citep{Selhorst2019}, the AR 12470 brightness temperature ($T_\mathrm{b}$) varies from 5800 to 6600~K across locations with magnetic field intensities greater than 1500~G, i.e., the minimum intensity to form a sunspot \citep{Livingston2012}. At 100~GHz, $T_{qS}=7110\pm90$~K \citep{Selhorst2019} and AR 12470 showed $T_\mathrm{b}$ varying from 7200 to 7650~K where $|\vec{B}|>1500$~G.

We also used 17 GHz radio maps obtained routinely by NoRH with resolution of 10-18 arcsec in intensity and circular polarization \citep{Nakajima1994} and intensity maps at 34~GHz with 5-10'' spatial resolution \citep{Takano1997}. We used a 17 GHz map observed at 02:44 UT and available in $512\times512$ pixels$^2$ format (4.91'' resolution), rotated longitudinally by $7.31^\circ$ with respect to the HMI magnetogram. At 17~GHz, $T_{qS}=10050\pm 120$~K, and AR 12470 has brightness temperatures varying from $5.3\times10^4$ to $6.8\times10^4$~K in the areas with $|\vec{B}|>1500$~G. Figure~\ref{fig:mapradio} shows the ALMA single dish maps obtained at 230~GHz and 100 GHz, plus the NoRH map at 17~GHz. In contrast to the small levels of variability in the ALMA maps, at 17~GHz the active region $T_\mathrm{b}$ reached values almost 7 times the level of the quiet Sun. The bright compact source in the active region has a high polarization degree ($\sim100$\%) indicating the presence of a gyroresonance core \citep{Vourlidas2006,Selhorst2008} which is coincident with the sunspot umbra.

\section{Results}\label{sec:analysis}

\begin{figure}[ht!]
\begin{center}
\includegraphics[width=9cm]{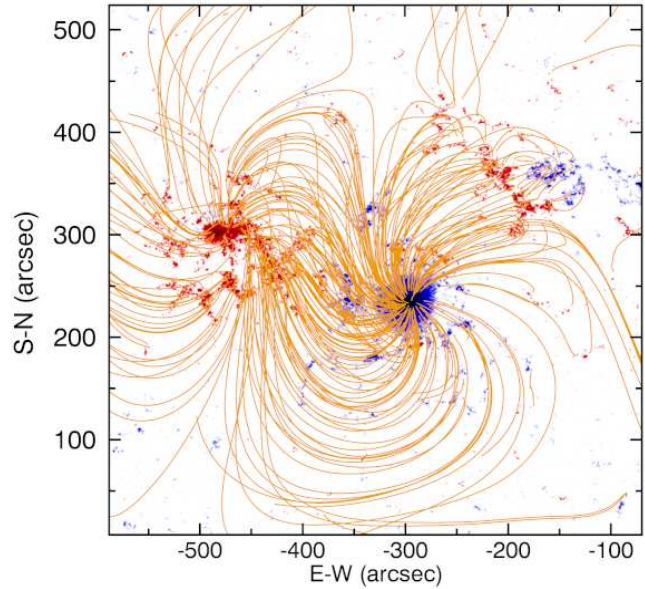}
\end{center}
\caption{Magnetic field lines obtained in a linear force-free extrapolation of the AR 12470 magnetogram, for $|\vec{B}|_{min}=200$~G and $\alpha=-0.005$.}
\label{fig:AIA}
\end{figure}

A low value for the $\alpha$ parameter in (1) does not significantly affect the extrapolated magnetic field configuration at low heights in the atmosphere: here $\alpha=-0.005$ was chosen for the LFF extrapolation since it generated a greater number of closed magnetic field lines in the selected region than did the potential-field extrapolation ($\alpha=0$). Figure~\ref{fig:AIA} shows some of the magnetic field lines obtained in the extrapolation with $\alpha=-0.005$ and $|\vec{B}|_{min}=200$~G. 

To speed up the computational process, the datacubes and images were resized from $1024\times1024$~pixels$^2$ to $256\times256$~pixels$^2$, corresponding to a spatial resolution $\sim2''$ per pixel. However, the radiative transfer calculations require better resolution in the height of the voxels ($h_z$) due to the relatively narrow width of the chromosphere, and they were re-scaled via interpolation from the resolution of the magnetogram to 50~km, similar to the resolution of the SSC model. 

The GA optimizes the electron density and temperature profiles for each vertical column in the model cube by comparing the brightness temperatures obtained from the model with the observed maps, pixel by pixel. This procedure initially generated $\sim$28~thousand different atmosphere profiles. These profiles were refined, in order to reduce the $\chi^2$ between observed $T_\mathrm{b}$ values and those obtained from the model. The final model is obtained when the  $\chi^2$ from the GA process ceases to reduce further.

\begin{figure}[ht!]
\begin{center}
\includegraphics[width=9cm]{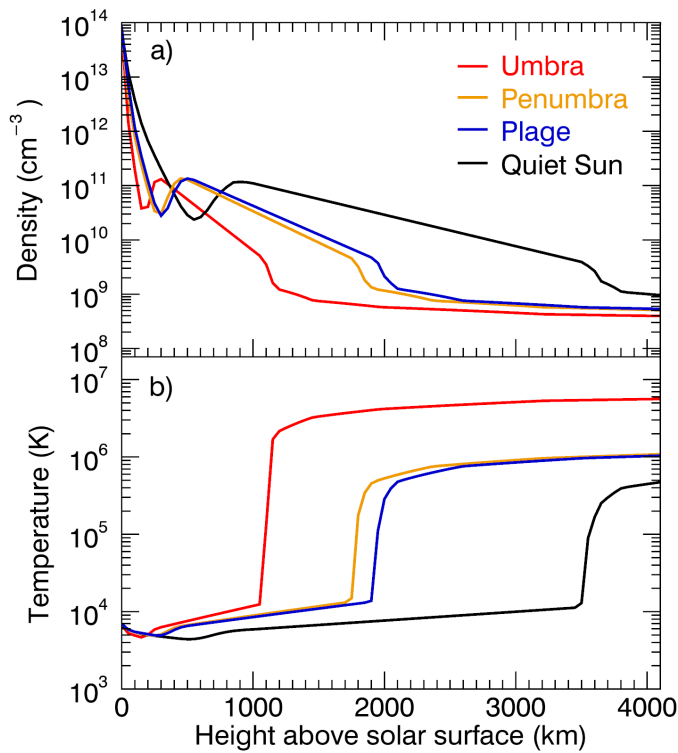}
\end{center}
\caption{Mean electron density (a) and temperature (b) profiles obtained for umbrae, penumbrae and plages. The SSC model (quiet Sun) is shown for comparison.}
\label{fig:temp_dens}
\end{figure}

The resulting atmospheric profiles were grouped into four different classes corresponding to different atmospheric features: umbrae, penumbrae, plage and quiet Sun. To speed up the GA process, distinct initial seeds were used for each class.  Each pixel in the magnetogram was classified as umbra, penumbra, or plage according to its intensity ($|\vec{B}|$), as follows: umbra, $|\vec{B}|\geq1500$~G ($\sim200$~pixels); penumbra, $1000\leq |\vec{B}_0|<1500$~G ($\sim$230~pixels); plages, $200\leq|\vec{B}_0|<1000$~G ($\sim$27200~pixels). All pixels with $|\vec{B}|\leq200$~G were classified as quiet Sun ($\sim$38000~pixels) and were assumed to have $T_B=T_{qS}$. We then obtained the average electron density and temperature profiles for each of the three AR classes by calculating the average of the profiles with the lowest $\chi^2$ values in the $T_\mathrm{b}$ calculation, in relation to the observed values at the three distinct observing frequencies. These resulting models are shown in Figure~\ref{fig:temp_dens} along with the the SSC quiet sun, for comparison. 

The $|\vec{B}|$ limits used to classify each pixel are based on studies by \cite{Livingston2012}, who derived a minimum necessary value of 1500~G for sunspot formation. Moreover, the umbra is a compact source with a diameter $\sim25''$ in its largest part. In \cite{Nita2018}, the authors proposed a similar separation, however, in that work the separation was based on solar white light images instead of a magnetogram, which is not useful for faint or spotless ARs. 

In the Table~\ref{Table_TR} we present the average height of the transition region (TR), which is one of the five free parameters fitted in the GA process, for each atmospheric class: thus, 1080$\pm$20~km above the solar surface for umbrae, 1800$\pm$50~km for penumbrae, and 2000$\pm$100~km for plages.

\begin{table*}[ht!]
\centering
\caption{Average transition region heights in models of distinct atmospheric features.} \label{Table_TR}
\begin{tabular}{lcccccc}
\hline
AR region     & $|\vec{B}|_{0}$~(G)  & TR height (km) \\ 
\hline
\hline
Quiet Sun    &   $|\vec{B}|_{0}<200$     & $3500$ \\ \hline 
Umbra        & $|\vec{B}|_{0}\geq1500$ & $1080 \pm~20$ \\ \hline 
Penumbra     & $1000\leq |\vec{B}_0|<1500$ & $1800 \pm~50$ \\ \hline 
Plages       & $200\leq|\vec{B}_0|<1000$ & $2000 \pm100$ \\ \hline 
\end{tabular}
\end{table*}

\begin{table*}[ht!]
\centering
\caption{Variation of temperature and density as function of height for the distinct AR areas. The distinct atmospheric layer were displayed in different colors, that is salmon for photosphere, green for chromosphere and orange for TR and corona.} \label{Table:profile}
\resizebox{\textwidth}{!}{
\begin{tabular}{rcc|cc|cc|cc}
\toprule
\hline
\multicolumn{9}{c}{}
\\ \hline \midrule
& \multicolumn{2}{c|}{Quiet Sun} & \multicolumn{2}{c|}{Plage} & \multicolumn{2}{c|}{Penumbra} & \multicolumn{2}{c}{Umbra} \\ \hline \midrule
\begin{tabular}[c|]{@{}c@{}}Height\\(km)\end{tabular} & \begin{tabular}[c|]{@{}c@{}}Temperature\\(K)\end{tabular} & \begin{tabular}[c|]{@{}c@{}}Density [$(n_e\times n_i)^{1/2}$]\\(cm$^{-3}$)\end{tabular} & \begin{tabular}[c]{@{}c@{}}Temperature\\(K)\end{tabular} & \begin{tabular}[c]{@{}c@{}}Density [$(n_e\times n_i)^{1/2}$]\\(cm$^{-3}$)\end{tabular} & \begin{tabular}[c]{@{}c@{}}Temperature\\(K)\end{tabular} & \begin{tabular}[c]{@{}c@{}}Density [$(n_e\times n_i)^{1/2}$]\\(cm$^{-3}$)\end{tabular} & \begin{tabular}[c]{@{}c@{}}Temperature\\(K)\end{tabular} & \begin{tabular}[c]{@{}c@{}}Density [$(n_e\times n_i)^{1/2}$]\\(cm$^{-3}$)\end{tabular} \\ \hline \midrule
$   0.0$ & \cellcolor[HTML]{FFCCC9}$ 6.5200E+03$ & \cellcolor[HTML]{FFCCC9}$ 6.7882E+13$ & \cellcolor[HTML]{FFCCC9}$ 6.9568E+03$ & \cellcolor[HTML]{FFCCC9}$ 7.9951E+13$ & \cellcolor[HTML]{FFCCC9}$ 6.9568E+03$ & \cellcolor[HTML]{FFCCC9}$ 8.0156E+13$ & \cellcolor[HTML]{FFCCC9}$ 6.6786E+03$ & \cellcolor[HTML]{FFCCC9}$ 8.0079E+13$ \\[2.5pt]
$ 100.0$ & \cellcolor[HTML]{FFCCC9}$ 5.4100E+03$ & \cellcolor[HTML]{FFCCC9}$ 3.7211E+12$ & \cellcolor[HTML]{FFCCC9}$ 5.4831E+03$ & \cellcolor[HTML]{FFCCC9}$ 1.1535E+12$ & \cellcolor[HTML]{FFCCC9}$ 5.4236E+03$ & \cellcolor[HTML]{FFCCC9}$ 7.8583E+11$ & \cellcolor[HTML]{FFCCC9}$ 4.9185E+03$ & \cellcolor[HTML]{FFCCC9}$ 1.9381E+11$ \\[2.5pt]
$ 200.0$ & \cellcolor[HTML]{FFCCC9}$ 4.9900E+03$ & \cellcolor[HTML]{FFCCC9}$ 6.5751E+11$ & \cellcolor[HTML]{FFCCC9}$ 5.1178E+03$ & \cellcolor[HTML]{FFCCC9}$ 1.2238E+11$ & \cellcolor[HTML]{FFCCC9}$ 5.0417E+03$ & \cellcolor[HTML]{FFCCC9}$ 8.2822E+10$ & \cellcolor[HTML]{9AFF99}$ 5.0368E+03$ & \cellcolor[HTML]{9AFF99}$ 4.0661E+10$ \\[2.5pt]
$ 300.0$ & \cellcolor[HTML]{FFCCC9}$ 4.7700E+03$ & \cellcolor[HTML]{FFCCC9}$ 2.0174E+11$ & \cellcolor[HTML]{FFCCC9}$ 4.9405E+03$ & \cellcolor[HTML]{FFCCC9}$ 2.7854E+10$ & \cellcolor[HTML]{9AFF99}$ 5.0807E+03$ & \cellcolor[HTML]{9AFF99}$ 3.0434E+10$ & \cellcolor[HTML]{9AFF99}$ 6.2980E+03$ & \cellcolor[HTML]{9AFF99}$ 1.3120E+11$ \\[2.5pt]
$ 400.0$ & \cellcolor[HTML]{FFCCC9}$ 4.5600E+03$ & \cellcolor[HTML]{FFCCC9}$ 6.9649E+10$ & \cellcolor[HTML]{9AFF99}$ 5.8022E+03$ & \cellcolor[HTML]{9AFF99}$ 7.6292E+10$ & \cellcolor[HTML]{9AFF99}$ 6.1542E+03$ & \cellcolor[HTML]{9AFF99}$ 1.0840E+11$ & \cellcolor[HTML]{9AFF99}$ 6.9067E+03$ & \cellcolor[HTML]{9AFF99}$ 8.5847E+10$ \\[2.5pt]
$ 500.0$ & \cellcolor[HTML]{FFCCC9}$ 4.4071E+03$ & \cellcolor[HTML]{FFCCC9}$ 2.8640E+10$ & \cellcolor[HTML]{9AFF99}$ 6.5441E+03$ & \cellcolor[HTML]{9AFF99}$ 1.3471E+11$ & \cellcolor[HTML]{9AFF99}$ 6.6928E+03$ & \cellcolor[HTML]{9AFF99}$ 1.2788E+11$ & \cellcolor[HTML]{9AFF99}$ 7.5747E+03$ & \cellcolor[HTML]{9AFF99}$ 5.5659E+10$ \\[2.5pt]
$ 600.0$ & \cellcolor[HTML]{9AFF99}$ 4.5500E+03$ & \cellcolor[HTML]{9AFF99}$ 2.6229E+10$ & \cellcolor[HTML]{9AFF99}$ 6.9024E+03$ & \cellcolor[HTML]{9AFF99}$ 1.1233E+11$ & \cellcolor[HTML]{9AFF99}$ 7.0932E+03$ & \cellcolor[HTML]{9AFF99}$ 9.8027E+10$ & \cellcolor[HTML]{9AFF99}$ 8.2983E+03$ & \cellcolor[HTML]{9AFF99}$ 3.6027E+10$ \\[2.5pt]
$ 700.0$ & \cellcolor[HTML]{9AFF99}$ 5.0045E+03$ & \cellcolor[HTML]{9AFF99}$ 5.2452E+10$ & \cellcolor[HTML]{9AFF99}$ 7.2778E+03$ & \cellcolor[HTML]{9AFF99}$ 8.8068E+10$ & \cellcolor[HTML]{9AFF99}$ 7.5130E+03$ & \cellcolor[HTML]{9AFF99}$ 7.5063E+10$ & \cellcolor[HTML]{9AFF99}$ 9.0839E+03$ & \cellcolor[HTML]{9AFF99}$ 2.3368E+10$ \\[2.5pt]
$ 800.0$ & \cellcolor[HTML]{9AFF99}$ 5.4690E+03$ & \cellcolor[HTML]{9AFF99}$ 9.4091E+10$ & \cellcolor[HTML]{9AFF99}$ 7.6690E+03$ & \cellcolor[HTML]{9AFF99}$ 6.8978E+10$ & \cellcolor[HTML]{9AFF99}$ 7.9529E+03$ & \cellcolor[HTML]{9AFF99}$ 5.7486E+10$ & \cellcolor[HTML]{9AFF99}$ 9.9380E+03$ & \cellcolor[HTML]{9AFF99}$ 1.5132E+10$ \\[2.5pt]
$ 900.0$ & \cellcolor[HTML]{9AFF99}$ 5.7445E+03$ & \cellcolor[HTML]{9AFF99}$ 1.1679E+11$ & \cellcolor[HTML]{9AFF99}$ 8.0776E+03$ & \cellcolor[HTML]{9AFF99}$ 5.4071E+10$ & \cellcolor[HTML]{9AFF99}$ 8.4148E+03$ & \cellcolor[HTML]{9AFF99}$ 4.4031E+10$ & \cellcolor[HTML]{FFC702}$ 1.0867E+04$ & \cellcolor[HTML]{FFC702}$ 9.8165E+09$ \\[2.5pt]
$1000.0$ & \cellcolor[HTML]{9AFF99}$ 5.9000E+03$ & \cellcolor[HTML]{9AFF99}$ 1.1000E+11$ & \cellcolor[HTML]{9AFF99}$ 8.5048E+03$ & \cellcolor[HTML]{9AFF99}$ 4.2397E+10$ & \cellcolor[HTML]{9AFF99}$ 8.9001E+03$ & \cellcolor[HTML]{9AFF99}$ 3.3727E+10$ & \cellcolor[HTML]{FFC702}$ 1.1879E+04$ & \cellcolor[HTML]{FFC702}$ 6.3579E+09$ \\[2.5pt]
$1100.0$ & \cellcolor[HTML]{9AFF99}$ 6.0573E+03$ & \cellcolor[HTML]{9AFF99}$ 9.6271E+10$ & \cellcolor[HTML]{9AFF99}$ 8.9519E+03$ & \cellcolor[HTML]{9AFF99}$ 3.3208E+10$ & \cellcolor[HTML]{9AFF99}$ 9.4106E+03$ & \cellcolor[HTML]{9AFF99}$ 2.5837E+10$ & \cellcolor[HTML]{FFC702}$ 1.3144E+05$ & \cellcolor[HTML]{FFC702}$ 3.5132E+09$ \\[2.5pt]
$1200.0$ & \cellcolor[HTML]{9AFF99}$ 6.2188E+03$ & \cellcolor[HTML]{9AFF99}$ 8.4253E+10$ & \cellcolor[HTML]{9AFF99}$ 9.4201E+03$ & \cellcolor[HTML]{9AFF99}$ 2.6047E+10$ & \cellcolor[HTML]{9AFF99}$ 9.9479E+03$ & \cellcolor[HTML]{9AFF99}$ 1.9794E+10$ & \cellcolor[HTML]{FFC702}$ 2.1610E+06$ & \cellcolor[HTML]{FFC702}$ 1.2177E+09$ \\[2.5pt]
$1300.0$ & \cellcolor[HTML]{9AFF99}$ 6.3847E+03$ & \cellcolor[HTML]{9AFF99}$ 7.3736E+10$ & \cellcolor[HTML]{9AFF99}$ 9.9106E+03$ & \cellcolor[HTML]{9AFF99}$ 2.0422E+10$ & \cellcolor[HTML]{FFC702}$ 1.0514E+04$ & \cellcolor[HTML]{FFC702}$ 1.5164E+10$ & \cellcolor[HTML]{FFC702}$ 2.5885E+06$ & \cellcolor[HTML]{FFC702}$ 1.0404E+09$ \\[2.5pt]
$1400.0$ & \cellcolor[HTML]{9AFF99}$ 6.5549E+03$ & \cellcolor[HTML]{9AFF99}$ 6.4532E+10$ & \cellcolor[HTML]{FFC702}$ 1.0425E+04$ & \cellcolor[HTML]{FFC702}$ 1.5996E+10$ & \cellcolor[HTML]{FFC702}$ 1.1109E+04$ & \cellcolor[HTML]{FFC702}$ 1.1618E+10$ & \cellcolor[HTML]{FFC702}$ 3.0181E+06$ & \cellcolor[HTML]{FFC702}$ 8.6324E+08$ \\[2.5pt]
$1500.0$ & \cellcolor[HTML]{9AFF99}$ 6.7297E+03$ & \cellcolor[HTML]{9AFF99}$ 5.6477E+10$ & \cellcolor[HTML]{FFC702}$ 1.0964E+04$ & \cellcolor[HTML]{FFC702}$ 1.2552E+10$ & \cellcolor[HTML]{FFC702}$ 1.1737E+04$ & \cellcolor[HTML]{FFC702}$ 8.9018E+09$ & \cellcolor[HTML]{FFC702}$ 3.3377E+06$ & \cellcolor[HTML]{FFC702}$ 7.4682E+08$ \\[2.5pt]
$1600.0$ & \cellcolor[HTML]{9AFF99}$ 6.9092E+03$ & \cellcolor[HTML]{9AFF99}$ 4.9427E+10$ & \cellcolor[HTML]{FFC702}$ 1.1529E+04$ & \cellcolor[HTML]{FFC702}$ 9.8406E+09$ & \cellcolor[HTML]{FFC702}$ 1.2398E+04$ & \cellcolor[HTML]{FFC702}$ 6.8206E+09$ & \cellcolor[HTML]{FFC702}$ 3.5101E+06$ & \cellcolor[HTML]{FFC702}$ 7.1085E+08$ \\[2.5pt]
$1700.0$ & \cellcolor[HTML]{9AFF99}$ 7.0934E+03$ & \cellcolor[HTML]{9AFF99}$ 4.3257E+10$ & \cellcolor[HTML]{FFC702}$ 1.2122E+04$ & \cellcolor[HTML]{FFC702}$ 7.7100E+09$ & \cellcolor[HTML]{FFC702}$ 1.3095E+04$ & \cellcolor[HTML]{FFC702}$ 5.2261E+09$ & \cellcolor[HTML]{FFC702}$ 3.6829E+06$ & \cellcolor[HTML]{FFC702}$ 6.7492E+08$ \\[2.5pt]
$1800.0$ & \cellcolor[HTML]{9AFF99}$ 7.2825E+03$ & \cellcolor[HTML]{9AFF99}$ 3.7857E+10$ & \cellcolor[HTML]{FFC702}$ 1.2744E+04$ & \cellcolor[HTML]{FFC702}$ 6.0504E+09$ & \cellcolor[HTML]{FFC702}$ 1.7611E+05$ & \cellcolor[HTML]{FFC702}$ 3.2130E+09$ & \cellcolor[HTML]{FFC702}$ 3.8559E+06$ & \cellcolor[HTML]{FFC702}$ 6.3902E+08$ \\[2.5pt]
$1900.0$ & \cellcolor[HTML]{9AFF99}$ 7.4767E+03$ & \cellcolor[HTML]{9AFF99}$ 3.3132E+10$ & \cellcolor[HTML]{FFC702}$ 1.3806E+04$ & \cellcolor[HTML]{FFC702}$ 4.7408E+09$ & \cellcolor[HTML]{FFC702}$ 4.5365E+05$ & \cellcolor[HTML]{FFC702}$ 1.3433E+09$ & \cellcolor[HTML]{FFC702}$ 4.0292E+06$ & \cellcolor[HTML]{FFC702}$ 6.0316E+08$ \\[2.5pt]
$2000.0$ & \cellcolor[HTML]{9AFF99}$ 7.6761E+03$ & \cellcolor[HTML]{9AFF99}$ 2.8996E+10$ & \cellcolor[HTML]{FFC702}$ 2.8696E+05$ & \cellcolor[HTML]{FFC702}$ 2.1365E+09$ & \cellcolor[HTML]{FFC702}$ 5.2845E+05$ & \cellcolor[HTML]{FFC702}$ 1.1619E+09$ & \cellcolor[HTML]{FFC702}$ 4.1805E+06$ & \cellcolor[HTML]{FFC702}$ 5.7397E+08$ \\[2.5pt]
$2100.0$ & \cellcolor[HTML]{9AFF99}$ 7.8808E+03$ & \cellcolor[HTML]{9AFF99}$ 2.5377E+10$ & \cellcolor[HTML]{FFC702}$ 4.7769E+05$ & \cellcolor[HTML]{FFC702}$ 1.2613E+09$ & \cellcolor[HTML]{FFC702}$ 5.8934E+05$ & \cellcolor[HTML]{FFC702}$ 1.0535E+09$ & \cellcolor[HTML]{FFC702}$ 4.2745E+06$ & \cellcolor[HTML]{FFC702}$ 5.6185E+08$ \\[2.5pt]
$2200.0$ & \cellcolor[HTML]{9AFF99}$ 8.0909E+03$ & \cellcolor[HTML]{9AFF99}$ 2.2209E+10$ & \cellcolor[HTML]{FFC702}$ 5.3609E+05$ & \cellcolor[HTML]{FFC702}$ 1.1487E+09$ & \cellcolor[HTML]{FFC702}$ 6.5039E+05$ & \cellcolor[HTML]{FFC702}$ 9.4510E+08$ & \cellcolor[HTML]{FFC702}$ 4.3686E+06$ & \cellcolor[HTML]{FFC702}$ 5.4974E+08$ \\[2.5pt]
$2300.0$ & \cellcolor[HTML]{9AFF99}$ 8.3067E+03$ & \cellcolor[HTML]{9AFF99}$ 1.9437E+10$ & \cellcolor[HTML]{FFC702}$ 5.9190E+05$ & \cellcolor[HTML]{FFC702}$ 1.0498E+09$ & \cellcolor[HTML]{FFC702}$ 7.1160E+05$ & \cellcolor[HTML]{FFC702}$ 8.3678E+08$ & \cellcolor[HTML]{FFC702}$ 4.4626E+06$ & \cellcolor[HTML]{FFC702}$ 5.3765E+08$ \\[2.5pt]
$2400.0$ & \cellcolor[HTML]{9AFF99}$ 8.5282E+03$ & \cellcolor[HTML]{9AFF99}$ 1.7010E+10$ & \cellcolor[HTML]{FFC702}$ 6.4785E+05$ & \cellcolor[HTML]{FFC702}$ 9.5113E+08$ & \cellcolor[HTML]{FFC702}$ 7.6257E+05$ & \cellcolor[HTML]{FFC702}$ 7.5280E+08$ & \cellcolor[HTML]{FFC702}$ 4.5566E+06$ & \cellcolor[HTML]{FFC702}$ 5.2557E+08$ \\[2.5pt]
$2500.0$ & \cellcolor[HTML]{9AFF99}$ 8.7556E+03$ & \cellcolor[HTML]{9AFF99}$ 1.4887E+10$ & \cellcolor[HTML]{FFC702}$ 7.0397E+05$ & \cellcolor[HTML]{FFC702}$ 8.5237E+08$ & \cellcolor[HTML]{FFC702}$ 7.8713E+05$ & \cellcolor[HTML]{FFC702}$ 7.3078E+08$ & \cellcolor[HTML]{FFC702}$ 4.6507E+06$ & \cellcolor[HTML]{FFC702}$ 5.1349E+08$ \\[2.5pt]
$2600.0$ & \cellcolor[HTML]{9AFF99}$ 8.9890E+03$ & \cellcolor[HTML]{9AFF99}$ 1.3029E+10$ & \cellcolor[HTML]{FFC702}$ 7.5755E+05$ & \cellcolor[HTML]{FFC702}$ 7.5985E+08$ & \cellcolor[HTML]{FFC702}$ 8.1172E+05$ & \cellcolor[HTML]{FFC702}$ 7.0878E+08$ & \cellcolor[HTML]{FFC702}$ 4.7448E+06$ & \cellcolor[HTML]{FFC702}$ 5.0143E+08$ \\[2.5pt]
$2700.0$ & \cellcolor[HTML]{9AFF99}$ 9.2287E+03$ & \cellcolor[HTML]{9AFF99}$ 1.1402E+10$ & \cellcolor[HTML]{FFC702}$ 7.8152E+05$ & \cellcolor[HTML]{FFC702}$ 7.3639E+08$ & \cellcolor[HTML]{FFC702}$ 8.3633E+05$ & \cellcolor[HTML]{FFC702}$ 6.8679E+08$ & \cellcolor[HTML]{FFC702}$ 4.8389E+06$ & \cellcolor[HTML]{FFC702}$ 4.8937E+08$ \\[2.5pt]
$2800.0$ & \cellcolor[HTML]{9AFF99}$ 9.4748E+03$ & \cellcolor[HTML]{9AFF99}$ 9.9791E+09$ & \cellcolor[HTML]{FFC702}$ 8.0406E+05$ & \cellcolor[HTML]{FFC702}$ 7.1633E+08$ & \cellcolor[HTML]{FFC702}$ 8.6098E+05$ & \cellcolor[HTML]{FFC702}$ 6.6481E+08$ & \cellcolor[HTML]{FFC702}$ 4.9330E+06$ & \cellcolor[HTML]{FFC702}$ 4.7733E+08$ \\[2.5pt]
$2900.0$ & \cellcolor[HTML]{9AFF99}$ 9.7275E+03$ & \cellcolor[HTML]{9AFF99}$ 8.7335E+09$ & \cellcolor[HTML]{FFC702}$ 8.2661E+05$ & \cellcolor[HTML]{FFC702}$ 6.9630E+08$ & \cellcolor[HTML]{FFC702}$ 8.8565E+05$ & \cellcolor[HTML]{FFC702}$ 6.4285E+08$ & \cellcolor[HTML]{FFC702}$ 5.0271E+06$ & \cellcolor[HTML]{FFC702}$ 4.6529E+08$ \\[2.5pt]
$3000.0$ & \cellcolor[HTML]{9AFF99}$ 9.9868E+03$ & \cellcolor[HTML]{9AFF99}$ 7.6433E+09$ & \cellcolor[HTML]{FFC702}$ 8.4920E+05$ & \cellcolor[HTML]{FFC702}$ 6.7627E+08$ & \cellcolor[HTML]{FFC702}$ 9.1035E+05$ & \cellcolor[HTML]{FFC702}$ 6.2090E+08$ & \cellcolor[HTML]{FFC702}$ 5.1213E+06$ & \cellcolor[HTML]{FFC702}$ 4.5325E+08$ \\[2.5pt]
$3100.0$ & \cellcolor[HTML]{FFC702}$ 1.0253E+04$ & \cellcolor[HTML]{FFC702}$ 6.6892E+09$ & \cellcolor[HTML]{FFC702}$ 8.7181E+05$ & \cellcolor[HTML]{FFC702}$ 6.5625E+08$ & \cellcolor[HTML]{FFC702}$ 9.3507E+05$ & \cellcolor[HTML]{FFC702}$ 5.9897E+08$ & \cellcolor[HTML]{FFC702}$ 5.2155E+06$ & \cellcolor[HTML]{FFC702}$ 4.4122E+08$ \\[2.5pt]
$3200.0$ & \cellcolor[HTML]{FFC702}$ 1.0527E+04$ & \cellcolor[HTML]{FFC702}$ 5.8542E+09$ & \cellcolor[HTML]{FFC702}$ 8.9443E+05$ & \cellcolor[HTML]{FFC702}$ 6.3625E+08$ & \cellcolor[HTML]{FFC702}$ 9.5982E+05$ & \cellcolor[HTML]{FFC702}$ 5.7704E+08$ & \cellcolor[HTML]{FFC702}$ 5.3098E+06$ & \cellcolor[HTML]{FFC702}$ 4.2920E+08$ \\[2.5pt]
$3300.0$ & \cellcolor[HTML]{FFC702}$ 1.0807E+04$ & \cellcolor[HTML]{FFC702}$ 5.1235E+09$ & \cellcolor[HTML]{FFC702}$ 9.1709E+05$ & \cellcolor[HTML]{FFC702}$ 6.1626E+08$ & \cellcolor[HTML]{FFC702}$ 9.7439E+05$ & \cellcolor[HTML]{FFC702}$ 5.6817E+08$ & \cellcolor[HTML]{FFC702}$ 5.3552E+06$ & \cellcolor[HTML]{FFC702}$ 4.2413E+08$ \\[2.5pt]
$3400.0$ & \cellcolor[HTML]{FFC702}$ 1.1095E+04$ & \cellcolor[HTML]{FFC702}$ 4.4839E+09$ & \cellcolor[HTML]{FFC702}$ 9.3976E+05$ & \cellcolor[HTML]{FFC702}$ 5.9627E+08$ & \cellcolor[HTML]{FFC702}$ 9.8782E+05$ & \cellcolor[HTML]{FFC702}$ 5.6076E+08$ & \cellcolor[HTML]{FFC702}$ 5.3896E+06$ & \cellcolor[HTML]{FFC702}$ 4.2060E+08$ \\[2.5pt]
$3500.0$ & \cellcolor[HTML]{FFC702}$ 1.2921E+04$ & \cellcolor[HTML]{FFC702}$ 3.9172E+09$ & \cellcolor[HTML]{FFC702}$ 9.6246E+05$ & \cellcolor[HTML]{FFC702}$ 5.7630E+08$ & \cellcolor[HTML]{FFC702}$ 1.0012E+06$ & \cellcolor[HTML]{FFC702}$ 5.5335E+08$ & \cellcolor[HTML]{FFC702}$ 5.4239E+06$ & \cellcolor[HTML]{FFC702}$ 4.1709E+08$ \\[2.5pt]
$3600.0$ & \cellcolor[HTML]{FFC702}$ 1.6723E+05$ & \cellcolor[HTML]{FFC702}$ 2.6584E+09$ & \cellcolor[HTML]{FFC702}$ 9.7625E+05$ & \cellcolor[HTML]{FFC702}$ 5.6767E+08$ & \cellcolor[HTML]{FFC702}$ 1.0147E+06$ & \cellcolor[HTML]{FFC702}$ 5.4595E+08$ & \cellcolor[HTML]{FFC702}$ 5.4581E+06$ & \cellcolor[HTML]{FFC702}$ 4.1357E+08$ \\[2.5pt]
$3700.0$ & \cellcolor[HTML]{FFC702}$ 2.9924E+05$ & \cellcolor[HTML]{FFC702}$ 1.5281E+09$ & \cellcolor[HTML]{FFC702}$ 9.8857E+05$ & \cellcolor[HTML]{FFC702}$ 5.6091E+08$ & \cellcolor[HTML]{FFC702}$ 1.0281E+06$ & \cellcolor[HTML]{FFC702}$ 5.3854E+08$ & \cellcolor[HTML]{FFC702}$ 5.4922E+06$ & \cellcolor[HTML]{FFC702}$ 4.1006E+08$ \\[2.5pt]
$3800.0$ & \cellcolor[HTML]{FFC702}$ 3.9164E+05$ & \cellcolor[HTML]{FFC702}$ 1.1247E+09$ & \cellcolor[HTML]{FFC702}$ 1.0009E+06$ & \cellcolor[HTML]{FFC702}$ 5.5416E+08$ & \cellcolor[HTML]{FFC702}$ 1.0415E+06$ & \cellcolor[HTML]{FFC702}$ 5.3115E+08$ & \cellcolor[HTML]{FFC702}$ 5.5263E+06$ & \cellcolor[HTML]{FFC702}$ 4.0655E+08$ \\[2.5pt]
$3900.0$ & \cellcolor[HTML]{FFC702}$ 4.2154E+05$ & \cellcolor[HTML]{FFC702}$ 1.0526E+09$ & \cellcolor[HTML]{FFC702}$ 1.0132E+06$ & \cellcolor[HTML]{FFC702}$ 5.4742E+08$ & \cellcolor[HTML]{FFC702}$ 1.0550E+06$ & \cellcolor[HTML]{FFC702}$ 5.2376E+08$ & \cellcolor[HTML]{FFC702}$ 5.5602E+06$ & \cellcolor[HTML]{FFC702}$ 4.0304E+08$ \\[2.5pt]
$4000.0$ & \cellcolor[HTML]{FFC702}$ 4.4674E+05$ & \cellcolor[HTML]{FFC702}$ 1.0055E+09$ & \cellcolor[HTML]{FFC702}$ 1.0255E+06$ & \cellcolor[HTML]{FFC702}$ 5.4067E+08$ & \cellcolor[HTML]{FFC702}$ 1.0684E+06$ & \cellcolor[HTML]{FFC702}$ 5.1637E+08$ & \cellcolor[HTML]{FFC702}$ 5.5941E+06$ & \cellcolor[HTML]{FFC702}$ 3.9953E+08$ \\[2.5pt]
$4250.0$ & \cellcolor[HTML]{FFC702}$ 5.0974E+05$ & \cellcolor[HTML]{FFC702}$ 8.8756E+08$ & \cellcolor[HTML]{FFC702}$ 1.0563E+06$ & \cellcolor[HTML]{FFC702}$ 5.2383E+08$ & \cellcolor[HTML]{FFC702}$ 1.1020E+06$ & \cellcolor[HTML]{FFC702}$ 4.9791E+08$ & \cellcolor[HTML]{FFC702}$ 5.6786E+06$ & \cellcolor[HTML]{FFC702}$ 3.9078E+08$ \\[2.5pt]
$4500.0$ & \cellcolor[HTML]{FFC702}$ 5.7275E+05$ & \cellcolor[HTML]{FFC702}$ 7.6965E+08$ & \cellcolor[HTML]{FFC702}$ 1.0871E+06$ & \cellcolor[HTML]{FFC702}$ 5.0701E+08$ & \cellcolor[HTML]{FFC702}$ 1.1356E+06$ & \cellcolor[HTML]{FFC702}$ 4.7948E+08$ & \cellcolor[HTML]{FFC702}$ 5.7628E+06$ & \cellcolor[HTML]{FFC702}$ 3.8203E+08$ \\[2.5pt]
$4750.0$ & \cellcolor[HTML]{FFC702}$ 6.3281E+05$ & \cellcolor[HTML]{FFC702}$ 6.5889E+08$ & \cellcolor[HTML]{FFC702}$ 1.1179E+06$ & \cellcolor[HTML]{FFC702}$ 4.9020E+08$ & \cellcolor[HTML]{FFC702}$ 1.1692E+06$ & \cellcolor[HTML]{FFC702}$ 4.6107E+08$ & \cellcolor[HTML]{FFC702}$ 5.8466E+06$ & \cellcolor[HTML]{FFC702}$ 3.7330E+08$ \\[2.5pt]
$5000.0$ & \cellcolor[HTML]{FFC702}$ 6.5724E+05$ & \cellcolor[HTML]{FFC702}$ 6.3504E+08$ & \cellcolor[HTML]{FFC702}$ 1.1488E+06$ & \cellcolor[HTML]{FFC702}$ 4.7341E+08$ & \cellcolor[HTML]{FFC702}$ 1.2028E+06$ & \cellcolor[HTML]{FFC702}$ 4.4267E+08$ & \cellcolor[HTML]{FFC702}$ 5.9301E+06$ & \cellcolor[HTML]{FFC702}$ 3.6457E+08$ \\[2.5pt]
$5250.0$ & \cellcolor[HTML]{FFC702}$ 6.8167E+05$ & \cellcolor[HTML]{FFC702}$ 6.1119E+08$ & \cellcolor[HTML]{FFC702}$ 1.1796E+06$ & \cellcolor[HTML]{FFC702}$ 4.5664E+08$ & \cellcolor[HTML]{FFC702}$ 1.2355E+06$ & \cellcolor[HTML]{FFC702}$ 4.2486E+08$ & \cellcolor[HTML]{FFC702}$ 6.0135E+06$ & \cellcolor[HTML]{FFC702}$ 3.5585E+08$ \\[2.5pt]
$5500.0$ & \cellcolor[HTML]{FFC702}$ 7.0610E+05$ & \cellcolor[HTML]{FFC702}$ 5.8734E+08$ & \cellcolor[HTML]{FFC702}$ 1.2104E+06$ & \cellcolor[HTML]{FFC702}$ 4.3988E+08$ & \cellcolor[HTML]{FFC702}$ 1.2479E+06$ & \cellcolor[HTML]{FFC702}$ 4.1947E+08$ & \cellcolor[HTML]{FFC702}$ 6.0966E+06$ & \cellcolor[HTML]{FFC702}$ 3.4714E+08$ \\[2.5pt]
$5750.0$ & \cellcolor[HTML]{FFC702}$ 7.3052E+05$ & \cellcolor[HTML]{FFC702}$ 5.6349E+08$ & \cellcolor[HTML]{FFC702}$ 1.2402E+06$ & \cellcolor[HTML]{FFC702}$ 4.2383E+08$ & \cellcolor[HTML]{FFC702}$ 1.2602E+06$ & \cellcolor[HTML]{FFC702}$ 4.1409E+08$ & \cellcolor[HTML]{FFC702}$ 6.1795E+06$ & \cellcolor[HTML]{FFC702}$ 3.3844E+08$ \\[2.5pt]
$6000.0$ & \cellcolor[HTML]{FFC702}$ 7.5495E+05$ & \cellcolor[HTML]{FFC702}$ 5.3964E+08$ & \cellcolor[HTML]{FFC702}$ 1.2515E+06$ & \cellcolor[HTML]{FFC702}$ 4.1892E+08$ & \cellcolor[HTML]{FFC702}$ 1.2724E+06$ & \cellcolor[HTML]{FFC702}$ 4.0872E+08$ & \cellcolor[HTML]{FFC702}$ 6.2623E+06$ & \cellcolor[HTML]{FFC702}$ 3.2975E+08$\\[2.5pt]
\hline
\bottomrule
\end{tabular}}
\end{table*}

\subsection{Free-free contribution}\label{sec:free-free}

The GA is able to derive appropriate fits for the assumed basic structure of the solar atmosphere, i.e., fitting for the 5 free parameters yields satisfactory models. {Although, the SSC model considers the plasma fully ionized above 1000 km, i.e. $n_e=n_i$, to simplify the GA the densities were grouped in a single variable, that is $(n_e\times n_i)^{1/2}$. This variable was assumed as the density in  Table~\ref{Table:profile} and  Figure~\ref{fig:temp_dens}. Table~\ref{Table:profile} shows the variation of temperature and density as function of height for the distinct AR areas, moreover, the distinct atmospheric layer were displayed in different colors, that is salmon for photosphere, green for chromosphere and orange for TR and corona.}   

Comparison with the quiet-Sun model indicates that the umbrae, penumbrae and plages have a narrower temperature-minimum region and a thinner chromosphere than the quiet Sun, as can be seen in {Table~\ref{Table:profile} and  Figure~\ref{fig:temp_dens}}. Moreover, the compression of the temperature minimum and the chromosphere are most pronounced in umbral pixels, i.e. the region with more intense magnetic fields (Figure~\ref{fig:temp_dens}). The mean umbral model placed the transition region close to 1000~km above the solar surface, in agreement with previous work \citep[e.g., ][see also Zlotnik et al., 1996]{Fontenla1999,Fontenla2009,Selhorst2008,Nita2018}. \nocite{Zlotnik1996} Furthermore, the umbral region is the only one that presents a significant increase in the coronal temperature, which is necessary to reach the high $T_\mathrm{b}$ values observed in the gyroresonance source at 17~GHz \citep{Selhorst2008,Selhorst2009}.

\begin{figure}[ht!] 
\begin{center}
\includegraphics[width=17cm]{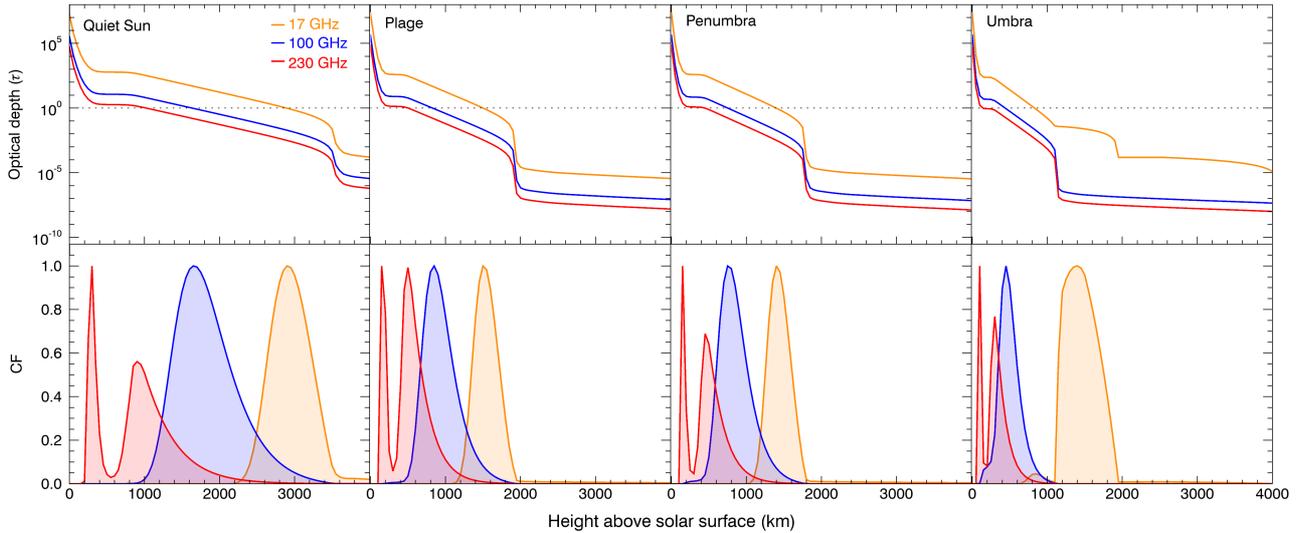}
\end{center}
   \caption{{\it Top}: the panels show the optical depth ($\tau$) variation with the height above the solar surface for the following areas: quiet Sun, plage, penumbra and umbra. The horizontal dotted line represents the $\tau=1$ height. {\it Bottom}: the variation of the contribution function (CF) with the height above the solar surface. Each CF was normalized by its maximum for a better comparison of the formation height of the emission at each frequency.}
   \label{fig:tau_cf}
\end{figure}

In Figure~\ref{fig:tau_cf} we show, for each atmospheric class, the optical depth ($\tau$, top row) and the contribution function (CF, bottom row), that represents the emission variation with the atmosphere height. The CF is defined as

\begin{equation}\label{func_cf}
{\rm CF}(h)=j_\nu e^{-\tau_\nu}
\end{equation}
\noindent  where $j_\nu=\kappa_\nu B_\nu(T)$ is the emission coefficient, and  $B_\nu(T)$ is the Planck function. Following \cite{Tapia2020}, each CF shown in Figure~\ref{fig:temp_dens} was normalized by its maximum, for a better comparison of the formation height of the emission at each frequency.

\begin{figure}[ht!] 
\begin{center}
\includegraphics[width=9cm]{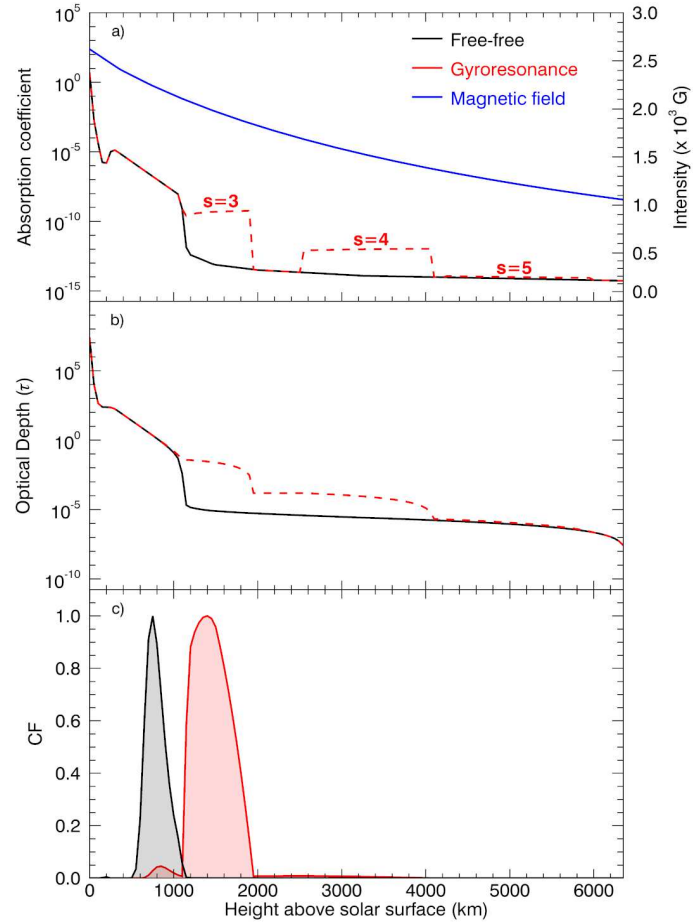} 
\end{center}
   \caption{a) Absorption coefficient $\kappa_\nu$ obtained at 17~GHz for the free-free (black curve) and the gyroresonance (red curve) for the highest modelled $T_\mathrm{b}$. The blue curve represents the magnetic field variation. b) Variation of the optical depth for both emission mechanisms. c) the variation of the contribution function (CF) with the height above the solar surface, in which the black curve represents the free-free emission,  and the red curve represents the total emission, i.e., free-free plus gyroresonance ($\kappa_b+\kappa_g$).}
   \label{fig:kappa17}
\end{figure}

As previously reported by \cite{Selhorst2019}, the quiet Sun free-free emission at 230~GHz is formed mainly in two different layers: near the temperature-minimum region ($h\sim400$~km), and in the chromosphere, with a peak around 900~km. Moreover, the contraction of the atmosphere does not change the CF double-peak structure, rather it just changes the heights of the two peaks and their contribution percentages (see red curves in Figure~\ref{fig:tau_cf}). On the other hand, the quiet Sun free-free emission at 100 and 17~GHz are completely formed in the chromosphere, with CF peaks, respectively, at 1650 and 2900~Km. These heights of the maximum of the CF are close to the atmosphere height at which $\tau=1$, as expected. Moreover, the atmospheric reduction of the AR active areas only moves the CF peaks closer to the solar surface. However, the 17~GHz $\tau$ results (orange curves) in the umbral region showed a huge change in the TR structure that cannot be attributed to free-free opacity and will be discussed in detail in the next section (\ref{sec:gyro}). The chromospheric free-free contribution is still present at 17 GHz, but its CF peak is only 5\% of the TR peak  (see the small orange peak close to 800~km).

\subsection{Gyroresonance contribution}\label{sec:gyro}

Several studies \citep[e.g. ][]{Shibasaki1994,White1997,Kundu2001,Vourlidas2006} have established that the gyroresonance radio emission is produced by opacity in harmonics 2, 3 and 4 of the electron cyclotron frequency, necessarily at TR or coronal heights in order to produce the brightness temperatures in excess of 10$^5$ K observed at 17~GHz. Moreover, \cite{Vourlidas2006} suggested that ARs with high polarization ($\gtrsim 30\%$) at 17 GHz have gyroresonance cores, and that $|\vec{B}|=2200$~G is the minimum intensity to generate such a core. AR 12470 presented a component with high polarization degree (up to $94\%$) and $|\vec{B}|\sim 2500$~G that is likely to be due to a gyroresonance source. 

To investigate the gyroresonance contribution, $\kappa_g$ was calculated for harmonics $s\leq 5$, allowing for an uncertainty of $10\%$ in the $|\vec{B}|$ values. That is, the 3rd harmonic was calculated where the magnetic field model showed intensities of $|\vec{B}|\simeq2000\pm200$~G, the 4th harmonic was calculated where $|\vec{B}|\simeq1500\pm150$~G, and the 5th harmonic where $|\vec{B}|\simeq1200\pm120$~G. Due to the heliographic position of AR 12740 (17.40$^\circ$N, 17.68$^\circ$E), it is estimated that $|\vec{B}|$ is 7.5\% greater than the observed $B_z$. 

Figure~\ref{fig:kappa17}a shows the variation in the absorption coefficient $\kappa_\nu$ with height along the line-of-sight to the location of maximum brightness temperature in AR 12470. The blue curve represents the magnetic field variation with height. The continuous black curve shows the absorption coefficient for bremsstrahlung ($\kappa_b$), and the dashed curve, in red, represents the contribution of gyroresonance ($\kappa_\nu=\kappa_b+\kappa_g$) and each of its respective harmonics ($s$). The 3rd harmonic occurs between 1150 and 1900~km in altitude, while the 4th harmonic occurs between 2550 and 4050 km. As for the 5th harmonic, it occurs between 4250 and 5950~km.

Figure~\ref{fig:kappa17}b shows the $\tau$ variation with the height, where the 3rd and 4th harmonics effectively contribute to $\tau$. The 3rd harmonic provides an optical depth of $\tau\sim10^{-1}$ (Figure~\ref{fig:kappa17}b red curve) that is three orders of magnitude greater than the free-free optical depth (Figure~\ref{fig:kappa17}b black curve). The contribution from the 4th harmonic is an order of magnitude larger than the free-free $\tau$. These results are in agreement with the studies of \cite{Shibasaki1994}, who found that the 3rd harmonic contribution is approximately three orders of magnitude greater than the 4th harmonic. Despite being optically thin, the gyroresonance contribution of $\tau\sim10^{-1}$ is able to increase the 17~GHz $T_\mathrm{b}$ from $10\times 10^3$~K to almost $70\times 10^3$~K thanks to the high temperature in the gyroresonance layers. To illustrate the contribution of each mechanism at 17 GHz, Figure~\ref{fig:kappa17}c shows the free-free CF in black and the total CF in red (free-free plus gyroresonance). While the free-free CF shows a single peak at around 800~km (black curve), when the gyroresonance is included an intense peak appears at TR/coronal heights (red curve) that is $\sim 20$~times greater than the free-free contribution.  

\subsection{Comparison of the synthetic and observed radio images}

\begin{figure}[ht!] 
\begin{center}
\includegraphics[width=13.5cm]{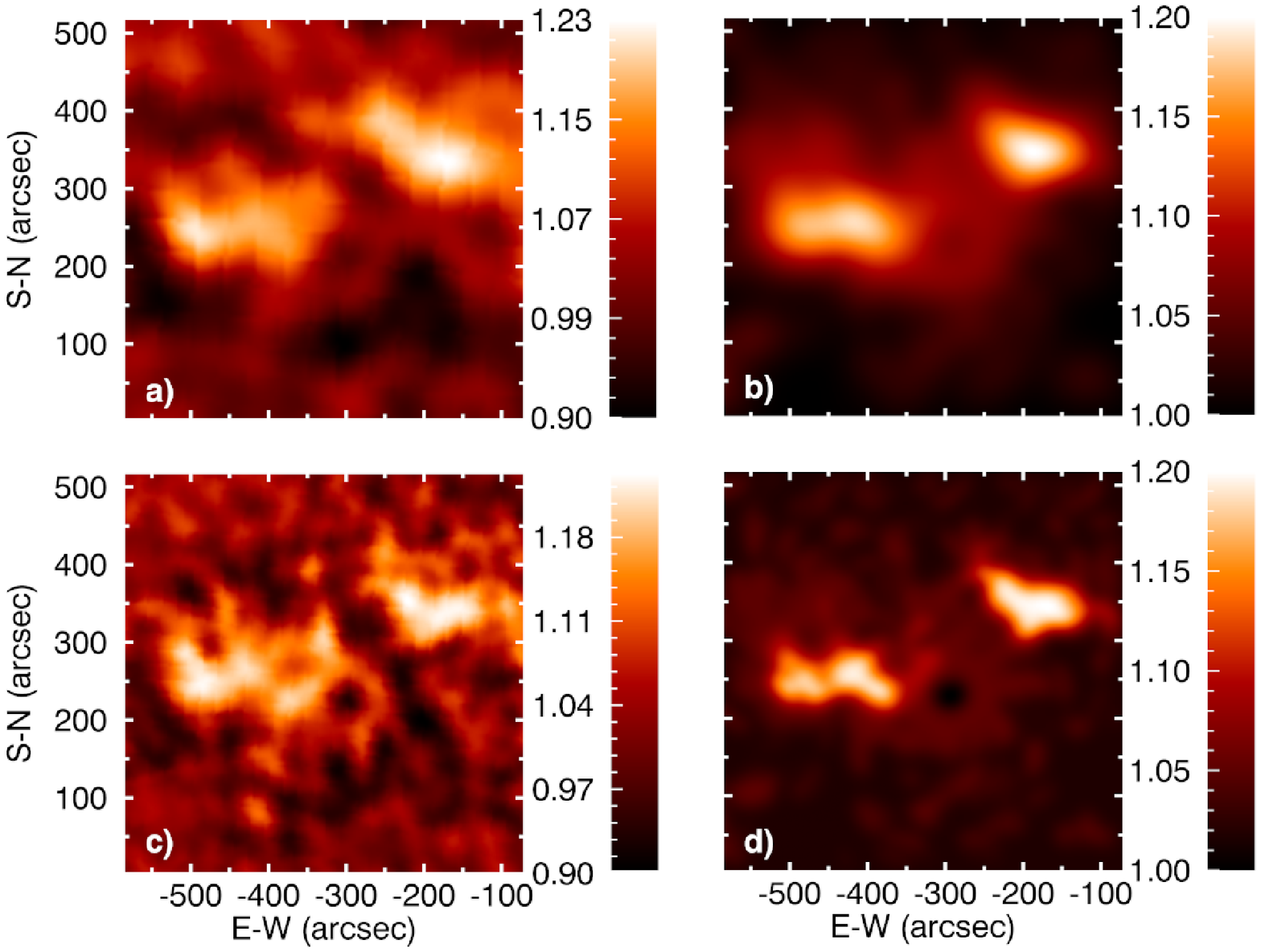}
\end{center}
   \caption{Comparison between the observations and the model results at ALMA wavelengths. The upper panels show the 100~GHz comparison between the (a) ALMA observation and the (b) model results convolved with a gaussian beam of  58'' resolution. The lower panels show (c) the ALMA observation at 230~GHz and (d) the model results convolved with a 25'' gaussian beam. As in Table~\ref{Table:mean}, the color bar scale is in multiples of the value of  $T_\mathrm{b,QS}$ appropriate for each frequency.}
   \label{fig:ALMA}

\begin{center}
\includegraphics[width=16.5cm]{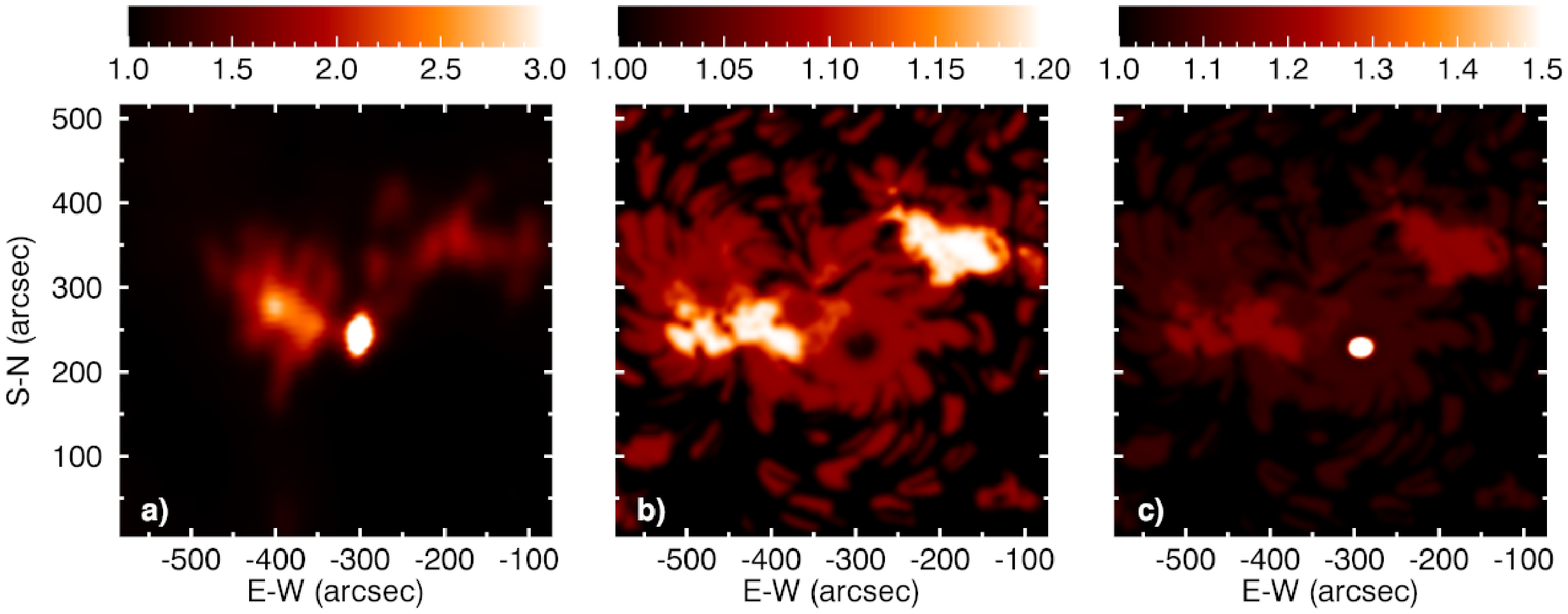}
\end{center}
   \caption{(a) NoRH observation of AR 12470 at 17~GHz, with the displayed intensities saturating at $3\times T_{qS}$~K (i.e., well below the peak $T_\mathrm{b}$ of the highly polarized source). (b) The GA model image at 17~GHz with 10'' resolution based only on the free-free opacity, and (c) the GA model result when both free-free and gyroresonance are included. As in Table~\ref{Table:TB17_pol}, the color bar scale is in multiples of  $T_\mathrm{b,QS}$ at 17 GHz (10000 K). }
   \label{fig:NoRH}
\end{figure}

The GA model images were generated with 2'' spatial resolution, which is much finer than the resolution of the radio observations. To compare the simulations with the observations, the model results were convolved with a 2D Gaussian beam that matches the resolution of the ALMA single-dish observations, i.e., 25'' and 58'' at 230 and 100~GHz, respectively.  In Figure~\ref{fig:ALMA} the 2D structure of AR 12470 observed with ALMA at 230 and 100~GHz (left panels) is compared with the model results (right panels). The upper row shows the 100~GHz comparison, while the lower row shows the 230 GHz comparison.

Due to better spatial resolution, the 230 GHz map shows more detail than that observed at 100~GHz, in both observation and model. The bright features observed at both frequencies are in good agreement with  the simulations. Moreover, due to the small size of the sunspot umbra (25'' diameter at its widest),  the dark umbral structure is readily seen at 230~GHz (see the center of panels \ref{fig:ALMA}c and \ref{fig:ALMA}d), but completely masked in the 100~GHz simulations due to the convolution with the bright structures around it (see Figure~\ref{fig:ALMA}a,b). 

Even though dark when compared with the surrounding areas, the umbral $T_\mathrm{b}$ is only smaller than the  $T_\mathrm{b,QS}$ in the 230~GHz observation, and in the model result without the beam convolution, as shown in Table~\ref{Table:mean}. When convolved to the observational resolution of 25'', the modelled umbral $T_\mathrm{b}$ at 230 GHz is at the same level as  $T_\mathrm{b,QS}$. Furthermore, as shown in Table~\ref{Table:mean}, the other AR areas (penumbra and plages) are brighter than the quiet Sun and showed a greater $T_\mathrm{b}$ variation than the umbra. All the simulated areas are consistent with the observations to within the uncertainties. 

\begin{table*}[ht!]
\centering
\caption{Comparison of the ALMA observations at 230 and 100~GHz and modeled averaged brightness temperatures of the distinct AR areas as compared with the quiet Sun temperature ( $T_\mathrm{b,QS}$).} \label{Table:mean}
\begin{tabular}{lcccccc}
\hline
AR region     & \multicolumn{5}{c}{Averaged Brightness Temperature ($\times T_{qS}$)}\\
\hline
& \multicolumn{3}{c}{230~GHz} & \multicolumn{3}{c}{100~GHz} \\
\hline
& Obs. & \multicolumn{2}{c}{Model} & Obs. & \multicolumn{2}{c}{Model}\\
\hline
& & 2'' & 25'' &  & 2'' & 58''\\
\hline
Umbra        & $0.99\pm0.04$ & $0.98\pm0.01$ & $1.00 \pm0.01$ & $1.03\pm0.01$ & $1.02\pm0.01$ & $1.06\pm0.01$  \\ \hline
Penumbra     & $1.05\pm0.05$ & $1.04\pm0.01$ & $1.02 \pm~0.02$ & $1.05\pm0.02$ & $1.07\pm0.01$ & $1.07\pm0.02$ \\ \hline
Plages       & $1.07\pm0.07$ & $1.08\pm0.05$ & $1.04 \pm0.04$ & $1.06\pm0.04$ & $1.10\pm0.04$ & $1.06\pm0.03$  \\ \hline
\end{tabular}
\end{table*}

The resulting low contrast of the dark umbra with respect to the quiet Sun in the model is due to the low resolution of the ALMA single-dish observations. Since the umbral size is almost equal to the ALMA 230~GHz beam, the umbral $T_\mathrm{b}$ is smoothed with the bright surrounding areas. At the much better resolution of the ALMA interferometric array, the AR 12470 umbra presented brightness temperature as low as $\sim 0.90\times T_{qS}$ \citep[see ][]{Shimojo2017}. Additionally, the 100~GHz ALMA interferometric observations obtained on December 16, 2015 showed a mysterious bright structure inside the umbra \citep{Iwai2017, Brajsa2021} that may still be present on December 17, increasing the umbral $T_\mathrm{b}$ and consequently the modeled value. 

Figure~\ref{fig:NoRH} compares the observation obtained at 17~GHz by NoRH (\ref{fig:NoRH}a) with the model results. Due to the high $T_\mathrm{b}$ in the polarized region ($\sim 7.0\times T_{qS}$), to visualize the other regions the image display range was saturated at $3.0\times T_{qS}$, allowing us to see bright free-free structures at positions consistent with those in the ALMA maps. Away from the polarized region, the maximum $T_\mathrm{b}$ was close to $2.8\times T_{qS}$, and the average value was $1.58\pm 0.32\times T_{qS}$ (see Table~\ref{Table:TB17_pol}). 

At 17~GHz the non-polarized regions appear in the model at positions matching the observed ones. However, the modeled atmospheric profiles that achieved good agreement with the $T_\mathrm{b}$ observations at 230 and 100~GHz were not able to reach the high $T_\mathrm{b}$ values observed at 17~GHz. While these regions showed observational values as high as $2.8\times T_{qS}$, the model was only able to reach a maximum $T_\mathrm{b}$ of $1.2\times T_{qS}$ (Figure~\ref{fig:NoRH}b). Moreover, since each modeled pixel used the same form of atmospheric profile to model the 3 frequencies (230, 100, and 17~GHz), we might expect the model to show good agreement with the bright and dark regions. The free-free regions bright at 17 GHz in the model (Figure~\ref{fig:NoRH}b) have good spatial agreement with those ones modeled at 230 and 100~GHz (Figure~\ref{fig:ALMA}b,d).   

If only  free-free emission is considered, at 17 GHz the umbra appears darker in the model than the surrounding area, with a brightness temperature of $1.06\pm0.02\times T_{qS}$ at 10'' resolution. When gyroresonance is included, the modeled $T_\mathrm{b}$ increased to values above $10 \times T_{qS}$. As summarized in Table~\ref{Table:TB17_pol},  when gyroresonance is taken into account the $T_\mathrm{b}$ values obtained for the highly polarized region are in agreement with those observed. Nevertheless, as can be observed in Figure~\ref{emiss_giro3}, the shape of the modelled polarized region is smaller and rounder, while the observed source appears elliptical. 

\begin{table}[]
\centering
\caption{Comparison of the 17~GHz observation and modeled averaged brightness temperatures of the polarized and non-polarized regions.}
\label{Table:TB17_pol}
\begin{tabular}{lccc|cc}
\hline
Bright region & \multicolumn{5}{c}{Averaged Brightness Temperature at 17~GHz ($\times T_{qS}$)}                             \\ \hline
              & \multicolumn{3}{c|}{$\kappa_b$}              & \multicolumn{2}{c}{$\kappa_b+\kappa_g$} \\ \hline
              & Obs.            & \multicolumn{2}{c|}{Model} & \multicolumn{2}{c}{Model}               \\ \hline
              &                 & 2''          & 10''        & 2''                 & 10'' \\ \hline
Polarized     & $3.69\pm 1.35$ & $1.06\pm 0.03$ & $1.06\pm 0.02$ & $9.41\pm 3.51$ & $4.08\pm1.19$ \\ \hline
Non-polarized & $1.58\pm 0.32$ & $1.04\pm 0.06$ & $1.04\pm 0.06$ & $1.04\pm 0.06$ & $1.04\pm 0.05$ \\ \hline
\end{tabular}
\end{table}

\section{Concluding remarks} \label{sec:discussion}    

 In this work, we present a data-constrained model of the solar atmosphere, in which we used the brightness temperatures of AR NOAA 12470 observed at three radio frequencies: 17~GHz from NoRH, and 100 and 230~GHz from ALMA single-dish data. Under the assumption that the radio emission originates from the combination of thermal free-free and gyroresonance processes, our model allows for calculating radio brightness temperature maps that can be compared with the observations. The magnetic field at distinct atmospheric heights was determined by a force-free field extrapolation using HMI/SDO photospheric magnetograms. In order to determine the best plasma temperature and density height profiles necessary to match the observations, the Pikaia genetic algorithm \citep{Charbonneau1995} is used to modify the standard quiet Sun atmospheric model characterized by 5 free parameters: the chromospheric gradients of temperature and electron density, the coronal temperature and density, and the TR height. The SSC \citep{Selhorst2005b} was used as the basic quiet Sun model, however, other models could be chosen as the basic model. The GA modified the SSC model to fit three distinct classes of active region features, defined by their magnetic field intensities: umbrae, penumbrae, and plages.

As seen in Figure~\ref{fig:ALMA} and Table~\ref{Table:mean}, at the ALMA wavelengths the model was in general agreement with the observations. The umbral region looks dark at 230~GHz and the brighter regions match the positions seen in the observations at both frequencies (230 and 100~GHz). However, due to the small size of the umbra (diameter $\lesssim 25''$), it is not apparent in the 100~GHz data or model. Moreover, as shown by the ALMA interferometric observations at 230~GHz \citep{Shimojo2017}, the umbral region is darker ($\sim 0.90\times T_{qS}$) than the single dish maps show ($0.99\pm0.04\times T_{qS}$), implying that the umbra will be darker at 100 and 17~GHz. Nevertheless, there are no observational data to confirm that. While at 17~GHz the umbral free-free emission is masked by the bright source due to gyroresonance opacity, at 100~GHz, the interferometric image showed the presence of a bright structure inside the umbral region \citep{Iwai2017}, that may be produced higher in the atmosphere than the region where the umbral emission is formed. 

Since the umbra is the region with the greatest magnetic field intensity ($|\vec{B}|$), it is also the location where the 17~GHz gyroresonance emission arises. As reported in previous works \citep{Shibasaki1994,Vourlidas2006,Selhorst2008}, gyroresonance emission at 17~GHz is typically formed at the 3rd harmonic ($|\vec{B}|=2000$~G), and needs to be well above the TR where the free-free emission is optically thin. However, the free-free contribution function  is much smaller than that of the gyroresonance (Figure~\ref{fig:kappa17}c). Moreover, if only thermal free-free emission was contributing opacity at 17~GHz, the AR umbra should look darker than the surrounding region (see Figure~\ref{fig:NoRH}b) and probably darker than the quiet Sun as the high resolution ALMA interferometric data suggest.

\section*{Funding}
This research was partially supported by FAPESP grant Nos.
2013/24155-3 and 2019/03301-8. PJAS acknowledges support from CNPq (contract 307612/2019-8).
SW was supported by the SolarALMA project, which has received funding from the European Research Council (ERC) under the European Union’s Horizon 2020 research and innovation programme (grant agreement No. 682462), and by the Research Council of Norway through its Centres of Excellence scheme, project number 262622.

\section*{Acknowledgments}

This paper makes use of the following ALMA data: ADS/JAO. ALMA\#2011.0.00020.SV. ALMA is a partnership of ESO (representing its member states), NSF (USA) and NINS (Japan), together with NRC (Canada) and NSC and ASIAA (Taiwan), and KASI (Republic of Korea), in co-operation with the Republic of Chile. The Joint ALMA Observatory is operated by ESO, AUI/NRAO, and NAOJ. The National Radio Astronomy Observatory is a facility of the National Science Foundation operated under cooperative agreement by Associated Universities, Inc. CGGC is grateful with FAPESP (2013/24155-3), CAPES (88887.310385/2018-00) and CNPq (307722/2019-8). 
RB acknowledges the support by the Croatian Science Foundation under the project 7549 “Millimeter and sub- millimeter observations of the solar chromosphere with ALMA.”

\bibliographystyle{frontiersinSCNS_ENG_HUMS} 
\bibliography{references}

\end{document}